# From Brain Imaging to Graph Analysis: a study on ADNI's patient cohort

Rui Zhang, PhD[1], Luca Giancardo[2], Danilo A. Pena[2], Yejin Kim[2],
Hanghang Tong[1], Xiaoqian Jiang[2]; for the Alzheimer's Disease Neuroimaging Initiative*
[1]School of Computing, Infor. & Decis. Syst. Engin., Arizona State University
[2]School of Biomedical Informatics, University of Texas Health Science Center at Houston

**ABSTRACT**
In this paper, we studied the association between the change of structural brain volumes to the potential development of Alzheimer's disease (AD). Using a simple abstraction technique, we converted regional cortical and subcortical volume differences over two time points for each study subject into a graph. We then obtained substructures of interest using a graph decomposition algorithm in order to extract pivotal nodes via multi-view feature selection. Intensive experiments using robust classification frameworks were conducted to evaluate the performance of using the brain substructures obtained under different thresholds. The results indicated that compact substructures acquired by examining the differences between patient groups were sufficient to discriminate between AD and healthy controls with an area under the receiver operating curve of 0.72.

**INTRODUCTION**
Brain functionality and decreasing cognition is known to be associated with the progression of Alzheimer's disease (AD). The difference between asymptomatic and symptomatic AD can change over time, and this period between being a healthy individual to having clinically present AD is referred to as mild cognitive impairment (MCI). This functional and cognitive decline is marked with memory lapses, poorer executive function, and increasing complexities associated with common activities of daily living that can last years [1]. The standard diagnosis of AD patients typically begins with a series of neuropsychological tests, clinical assessments, followed by various imaging tests. Over time, patients undergo multiple brain imaging visits at different points which allow physicians to link physical manifestations of AD to lab and clinical measurement data. This wealth of information enables researchers to take advantage of these multi-visit data to look at the AD neurodegeneration process (through observing cognitive decline) both cross-sectionally and longitudinally. Recent studies have used many methods look at AD from a different perspective. Researchers use techniques such as hierarchical classification [2], convolutional neural networks [3], tract-based spatial statistics [4], in addition to a whole series of multi-modality data [5] to improve AD diagnostic performance. Further, studies have looked at various combinations of phenotype classification, typically incorporating both stable and converted MCI patients [6]. This allows for a rich understanding of AD progression and of biomarkers that may be readily available to drive an improved and quicker AD diagnosis. In this work, we aim to leverage the advantages of graph-based approaches while discarding some of the disadvantages that come with diffusion MRI based techniques. Specifically, we used structural volume data to uncover how different regions of the brain are inter-connected during AD progression. We also incorporated longitudinal information which accounts for structural changes in a patient's brain over time. These inter-region network effects will be used for phenotype classification to demonstrate their discrimination power.

**RELATED WORK**
Graph theory and graph-based approaches have been studied for many years in the neuroimaging community. These networks have been created using several different modalities including structural magnetic resonance imaging (MRI), diffusion tensor imaging (DTI), and positron emission tomography (PET) for various diseases such as AD. Typically, these approaches look at the differences between normal controls (CN) and those with the studied disease. The subsequent analysis allows for interpretation of nodes, hubs, and edges that are important for characterizing these brain networks. The most relevant work to this study are those that use diffusion weighted MRI images (DWI) to construct AD brain networks. Studies have used DWI to understand white matter integrity differences in AD patients from graph statistics like fractional anisotropy with dimensionality reduction methods like independent component analysis [7]. These studies are motivated by the idea that AD is a disconnection-based disease [8]. Others have combined these techniques with amyloid pathology in preclinical AD patients. However, these authors find that this type of data are not sufficient to disrupt network metrics [9]. Many of these diffusion weighted techniques have commonly reported problems such as generating a considerable amount of false positive connections [10]. Further, the magnitude of the node and edge importance can differ between several techniques and methods used at the creation of the graph [11]. Researchers using graph analysis from structural brain networks uncovered different findings when looking at a large scale network versus several smaller sub-networks. Sub-networks showed that small worldness decreased in areas of the medial temporal lobe [10]. Thus, there is a need to develop alternative methods that utilize existing pipelines that have reliability and reproducibility. In addition, the aforementioned methods do not take into consideration longitudinal change over time from an individual level. With our novel methodology, we combined graph-based analysis, individual longitudinal changes, and phenotype classification to address some of these gaps.

*Data used in preparation of this article were obtained from the Alzheimer's Disease Neuroimaging Initiative (ADNI) database (adni.loni.usc.edu). As such, the investigators within the ADNI contributed to the design and implementation of ADNI and/or provided data but did not participate in analysis or writing of this report. A complete listing of ADNI investigators can be found at:
http://adni.loni.usc.edu/wp-content/uploads/how_to_apply/ADNI_Acknowledgement_List.pdf

**METHODS**
Data used in the preparation of this article were obtained from the Alzheimer's Disease Neuroimaging Initiative (ADNI) database (adni.loni.usc.edu). The ADNI was launched in 2003 as a public-private partnership, led by Principal Investigator Michael W. Weiner, MD. The primary goal of ADNI has been to test whether serial magnetic resonance imaging (MRI), positron emission tomography (PET), other biological markers, and clinical and neuropsychological assessment can be combined to measure the progression of mild cognitive impairment (MCI) and early Alzheimer's disease (AD).

We developed a method pipeline which includes image processing to extract the volume of different brain regions, multi-view learning based graph decomposition to identify critical pivotal points and subgraph structure, and meta-learning based robust classification. The goal is to demonstrate that the simple volumetric change of different regions in the brain can serve as a tool to discriminate AD, MCI and CN. The key is to find the relevant subgraph structures that can drive good model performance.

**Brain Image processing and differential brain imaging graph**
The FreeSurfer Longitudinal pipeline v 6.0 [12] was used to compute the cortical and subcortical region volumes used in this study [13]. This specific pipeline was designed to reduce the intra-subject variation in scans by registering each subject to an individualized common space. This template space takes into account all the available scans that patient has to reduce the global noise when comparing a specific subject's multiple scans at different time points as seen in Figure 1.

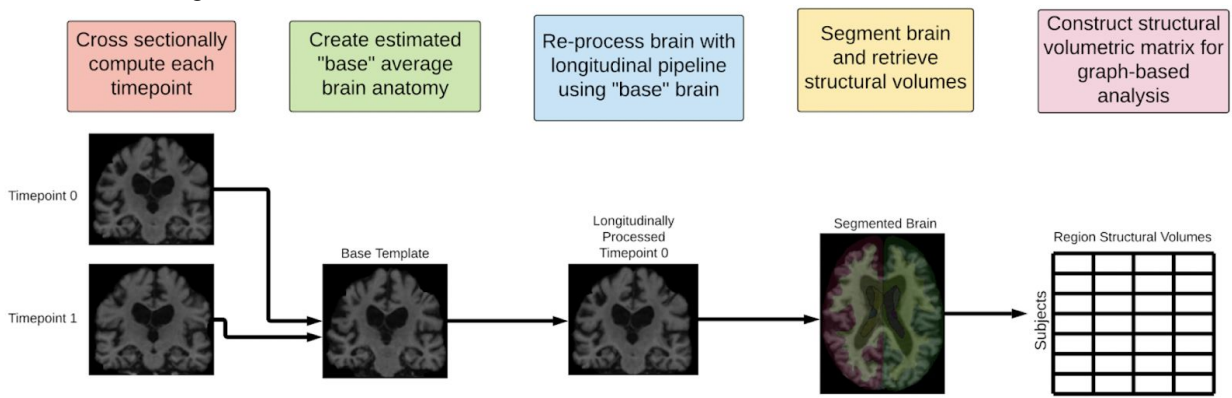

**Figure 1**: High-Level FreeSurfer Longitudinal Processing Pipeline

The pipeline includes common commands like motion correction and skull stripping. These images are then registered to the Talairach space to reduce voxel intensity noise within each subject. Brainmasks are then created followed by normalization, subcortical segmentation, surface reconstruction and cortical parcellation. Subsequently, segmented volume statistics are calculated. For this study's analysis, the statistics from "aseg.stats" and "aparc.stats" files output from the FreeSurfer pipeline were used. These files had an emphasis on white matter volume of different brain substructures. In total, there are 110 different volumes considered, and the subjects that were used each had two imaging timepoints (T0 and T1). This incorporates a richer set of information that encapsulates the neuro-related progression which occurs in AD and MCI patients. Because every patient is different, we modelled the change in volume of different brain regions to give an improved calibrated measurement of potential neurodegeneration. For every region of each patient $v$ (the size is denoted as $\eta_{vj}$), we computed their T1-T0 ratio change as $r_{vj} = \frac{\eta_{vj}^{T1} - \eta_{vj}^{T0}}{\eta_{vj}^{T0}}$. This ratio can be positive or negative as the brain region can shrink or enlarge over time. Then, we calculated $S^v_{jk} = r_{vj} * r_{vk}$ for each brain region pair for each patient to indicate if the regions are in concordance in terms of the directionality and intensity of change. If two regions stay unchanged, their product will be close to zero. Otherwise, $S^v_{jk}$ will have a bigger value, and the sign will correspond to the relative direction of each region, which can be positive or negative. We call the union of $S^v_{jk}$ (for every pairs $j$ and $k$) the differential brain imaging for patient $v$. We hypothesized that this metric could be related to underlying mechanisms of brain connectivity. Note that it is common that neighboring brain regions can show an opposite trend if one region shrinks quickly. This is due to the fact that neighboring regions will need to shrink in order to accommodate the enlargement of closeby regions. Figure 2 shows an example of our differential brain imaging graph as well as the histogram of its values. As one can tell, the values are very small, showing a strong sparsity in the structure.

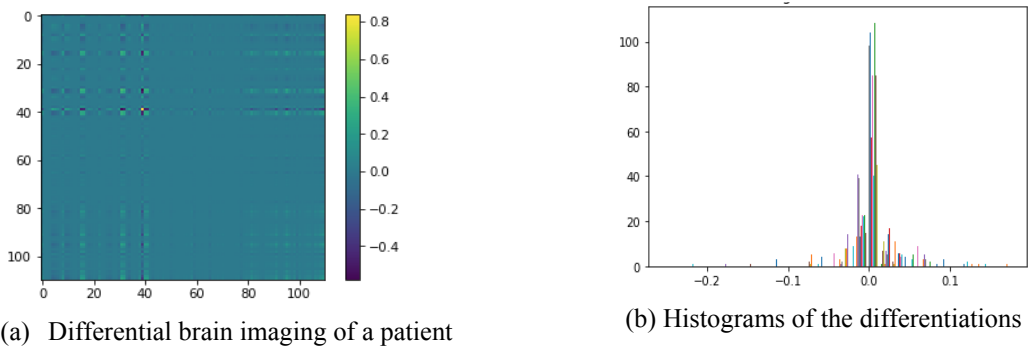

(a) Differential brain imaging of a patient    (b) Histograms of the differentiations

**Figure 2:** Illustration of the differential brain imaging used in this experiment.

**Multi-view model for graph decomposition**
*Pivotal nodes via multi-view feature selection*
In this subsection, a novel multi-view feature selection (MFS) technique was developed to mine the potential causative nodes of brain network for the Alzheimer's Disease Neuroimaging Initiative (ADNI) dataset. This method extracts a subgraph so that the pivotal nodes can be determined. Different from the conventional technique in which the similarity is frequently vectorized as input and redundant entries are involved, the proposed method is much more efficient because it selects the pivotal nodes instead of the entries of similarity. Under this mechanism, we only needed to pick out the informative feature subset of $d$ dimensional feature set instead of $d^2$ dimensional set, while avoiding the redundant entries of the symmetry similarity. Denote the similarity $S^{(v)} = [S^{(v)}_{ij}]_{d \times d} \in R^{d \times d}$ as the node graph regarding the brain network of each patient $v$ and the subspace $W \in R^{d \times k}$ as the selection matrix. If each patient is treated as a view, then the ranking of the potential causative nodes can be determined by solving the following graph-based model:

$$min_{W^T W = I} \sum_{v=1}^{l} \alpha_v \sum_{i=1}^{d} \sum_{j=1}^{d} S^{(v)}_{ij} \|W^i - W^j\|_2^2 + \lambda \|W\|_{2,1}, \quad (1)$$

where $\alpha_v$ is the prior weight assigned to the $v$-th view similarity, $W^i$ is the $i$-th row of the selection matrix $W$, and $\lambda$ is the regularization parameter to serve as a tradeoff between the error and the sparsity (controlling the robustness). More specifically, the importance or contribution of the $i$-th node can be evaluated by the $l_2$-norm of the $i$-th row of the selection matrix $W$ as $\|W^i\|_2$. Since the $l_{2,1}$ regularization in Eqn. (1), i.e. $\|W\|_{2,1} = \sum_{i=1}^{d} \|W^i\|_2$ can not be solved directly, we utilize the re-weighted method, such that the dual problem of Eqn. (1) can be formulated in the matrix form as

$$min_{W^T W = I} \sum_{v=1}^{l} \alpha_v Tr(W^T L^{(v)} W) + \lambda Tr(W^T D W), \quad (2)$$

where the Laplacian matrix of the $v$-th view is $L^{(v)} = G^{(v)} - S^{(v)} \in R^{d \times d}$ and degree matrix $G^{(v)} \in R^{d \times d}$ is diagonal with its $i$-th diagonal entry $G_{ii}^{(v)} = \sum_{j=1}^{d} S_{ij}^{(v)}$. Moreover, the diagonal matrix $D \in R^{d \times d}$ in Eqn. (2) serves as the transitional weight with updating its $i$-th diagonal entry $D_{ii} \leftarrow \frac{1}{2\sqrt{\|W^i\|_2^2 + \varepsilon}}$ until the convergence, where $\varepsilon$ is a very small perturbation number to avoid the trivial diagonal entry. To further solve problem (2), Lagrangian function can be represented:

$$L(W) = \sum_{v=1}^{l} \alpha_v Tr(W^T L^{(v)} W) + \lambda Tr(W^T D W) - Tr(\Lambda(W^T W - I)), \quad (3)$$

which leads to the KKT condition as $\frac{dL(W)}{dW} = 0 \Rightarrow (\sum_{v=1}^{l} \alpha_v L^{(v)} + \lambda D)W = W\Lambda$. In other words, $W$ can be easily solved by eigen-decomposition of the matrix $\sum_{v=1}^{l} \alpha_v L^{(v)} + \lambda D$, i.e., $W$ corresponds to the eigenvectors of the top $k$ smallest eigenvalues of $\sum_{v=1}^{l} \alpha_v L^{(v)} + \lambda D$. Eventually, the pseudo code can be summarized in Algorithm 1 to solve problem (2).

**Algorithm 1:** Pivotal node selection by solving the MFS dual in Eqn. (2)

---
Input: Similarity $S^{(v)} \in R^{d \times d}$ of each patient $v$.
Output: Node ranking.
  1: Initialize $D = I$ and Laplacian matrix $L^{(v)} = G^{(v)} - S^{(v)} \in R^{d \times d}$;
While **not converge**
  2: Update: $W \in R^{d \times k} \leftarrow argmin_{W^T W = I} Tr(W^T (\sum_{v=1}^{l} \alpha_v L^{(v)} + \lambda D)W)$;
   For $i = 1 : d$
  3: Update: $D_{ii} \leftarrow \frac{1}{2\sqrt{\|W^i\|_2^2 + \varepsilon}}$;
   End
End
  4: Sort node scores $\|W^i\|_2$.
---

*Representative/discriminative subgraph extraction*
The output from the previous pipeline provides critical ranking of the nodes based on their influence measured by the $l_2$-norm of the selection matrix (i.e., eigenvector matrix in a low dimensional projection of the Laplacian) in the similarity graph. The subgraph induced by these pivotal nodes contain critical information to represent individual cohorts which allows for characterizing the differences between different cohorts. The goal here is to extract robust structures that are representative and compact enough to discriminate the three groups (AD, MCI, CN). In particular, we are interested in: (1) subgraphs induced by the influential nodes for each group, and (2) subgraphs induced by the most influential nodes for pairwise graph subtraction (AD-MCI, AD-CN, MCI-CN). To decide the appropriate number of nodes, we set a pair of thresholds: K and r. These parameters stands for the top K ranked nodes for each parameters setting (tradeoff parameter $\lambda$ in Eq. 1 and the low dimensional projection parameter $k$) and the ratio $r$ to be consistently observed across all settings. These two parameters, which control the robustness of the pivotal points, get selected for downstream analysis.

**Classification using extracted subgraph structure**
In order to access the discrimination capacity of the identified pivotal nodes and subgraphs, we evaluated their separation power on the three groups of interest using classification models. Instead of picking a few fixed models for evaluation, we would like to ensure the robustness of our study. Different machine learning pipelines depend on different pre-processing steps (e.g., balancing, encoding, rescaling, etc.) and various parameter settings. Even experienced researchers might deploy sub-optimal (under limited time) models in the process of searching the best parameter combination in the complicated hyper-parameter space. This ultimately might leads to unreliable model evaluation and comparisons. On the other hand, it is not efficient to search the entire hyper-parameter space in a blind manner, especially when models are very expensive to compute. The recent advances in Bayesian optimization, meta-learning, and ensemble construction have led to some integrated solutions for both algorithm selection and hyperparameter tuning. Among the most popular ones, Auto-Sklearn [14] and Auto-Keras [15] represent state-of-the-art academic framework based on large scale ensembling and deep learning principles. Auto-Sklearn is used to create the best ensemble model from several base models. This abstracts the the modelling problem to a Combined Algorithm Selection and Hyperparameter optimization (CASH) to find the joint algorithms and hyperparameters together to minimize the learning objective. Auto-Sklearn uses a random forest driven sequential model-based optimization [16] to conduct general algorithm configuration (SMAC). Auto-Keras is used to automate the neural architecture search in designing better deep neural networks for the given task. In the heart of this framework, our method links network morphism with Bayesian optimization using neural network kernels for Gaussian process [17] and acquisition function optimization for tree structured space. Because these architectures are highly generalizable and comprehensive, we adopted them to train and test on our identified pivotal points and associated subgraph structures. Lastly, area under the receiver operating curve (AUC) scores were computed to evaluate the classification algorithms.

**EXPERIMENTS**
**Dataset:** The T1-weighted MRI images were downloaded from the ADNI website [18] in November 2018. The ADNI database is a collection of an array of imaging modalities of people at different stages of AD and at many sites, and it was created to allow researchers to uncover new findings about the neurodegenerative disease. In this study, we used only subjects with at least two separate imaging sessions, for a total of 857 patients. When more than two imaging sessions were identified, we selected the first and last. The first session was named "T0" the last "T1". Table 1 summarizes our study cohort in terms of demographics.

**Table 1:** Summarization of the demographics of the study subjects. Note that AD= Alzheimer's Disease, MCI= Mild Cognitive Impairment (subject self reporting symptoms, who may or may not develop AD), CN=Control normals (healthy subjects).

|  | **AD** | **MCI** | **CN** |
|---|---|---|---|
| Number of Patients | 213 | 322 | 322 |
| Age, years (mean, (s.d.)) | 75.6 (7.2) | 76.5 (7.6) | 77.4 (6.5) |
| Male gender (n (%)) | 108 (50.7%) | 203 (63%) | 164 (50.9%) |
| Race/Ethnicity - American Indian/Alaskan Native, Asian, Native Hawaiian/Other Pacific Islander, Black/African American White, More than one race, Unknown Not Hispanic or Latino (n (%)) | 0 (0%), 4 (1.9%), 0 (0%), 12 (5.6%), 194 (91.1%), 3 (1.4%) | 0 (0%), 8 (2.5%), 0 (0%), 12 (3.7%), 302 (93.8%), 0 (0%) | 0 (0%), 2 (1.9%), 0 (0%), 26 (8.1%), 288 (89.4%), 2 (0.6%) |

**Results**
As mentioned earlier, each patient have two MRI images T0 and T1, taken at different times. After image processing, the differences between these two time points were abstracted to 110 brain regions (nodes). These nodes were used to form the differential brain image graph of size 110x110. There were some measurement errors during image processing, which rendered some of these nodes invalid. Lastly, 101 nodes were used that were valid across all three groups (AD, MCI, CN). We conducted a comprehensive graph structural exploration and a grid search for the following hyperparameter space for tradeoff parameters $\lambda \in \{0.01, 0.1, 1, 10, 100\}$ in Eqn. 1 and low dimensional projection parameter $k \in \{15, 20, 25, 30, 35, 40, 45, 50, 55\}$. As for the graph subtraction, the class information were explicitly utilized to assign the different view weights $\alpha_v$ to formulate the graph subtraction between each two classes. For example, as to AD-MCI graph subtraction, $\alpha_v = \frac{322}{213}$ for AD, while $\alpha_v = -1$ for MCI. A total of 45 combinations of parameters ($\lambda$, $k$) were explored, and a passing criterion was defined as more than 40 times to be considered as a robust feature (pivotal node). We then identified pivotal nodes for the following three settings:

*Setting 1:* weighted graph subtraction between groups AD-MCI, AD-CN, and MCI-CN
*Setting 2:* AD, MCI, CN as individual groups
*Setting 3:* the entire cohort containing all three groups

The nodes identified in *setting 3* is close to the union of nodes in *setting 2* as data are relatively balanced. Data are randomly partitioned into 80% and 20% for training and testing. We did not conduct cross-validation because both Auto-Sklearn and Auto-Keras required significant resources but results were consistent with multiple runs.

(a) Auto-Sklearn (22 nodes): Union of top 20 pivotal nodes in ['am=(322/213)ad-mci', 'ac=(322/213)ad-cn','mc=mci-cn']

(b) Auto-Keras (22 nodes): Union of top 20 pivotal nodes in ['am=(322/213)ad-mci', 'ac=(322/213)ad-cn','mc=mci-cn']

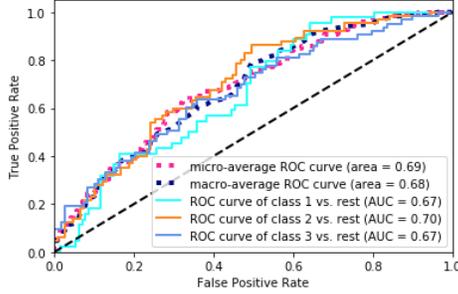
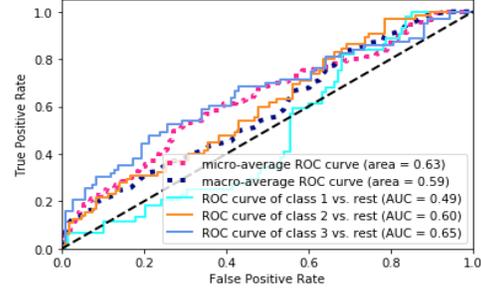

(c) Auto-Sklearn (40 nodes): Union of top 30 pivotal nodes in ['am=(322/213)ad-mci', 'ac=(322/213)ad-cn','mc=mci-cn']

(d) Auto-Keras (40 nodes): Union of top 30 pivotal nodes in ['am=(322/213)ad-mci', 'ac=(322/213)ad-cn','mc=mci-cn']

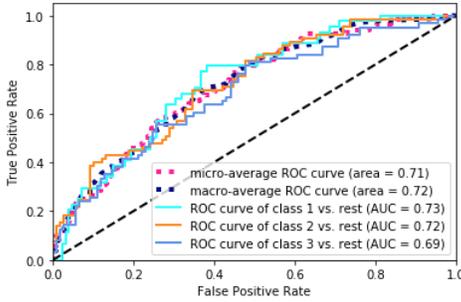
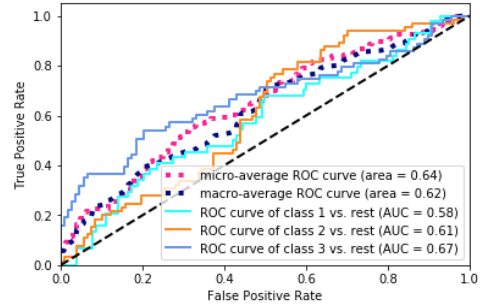

(e) Auto-Sklearn (77 nodes): Union of top 50 pivotal nodes in ['am=(322/213)ad-mci', 'ac=(322/213)ad-cn','mc=mci-cn']

(f) Auto-Keras (77 nodes): Union of top 50 pivotal nodes in ['am=(322/213)ad-mci', 'ac=(322/213)ad-cn','mc=mci-cn']

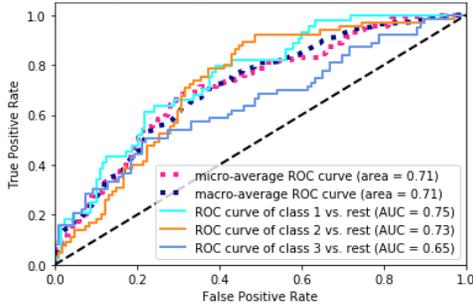
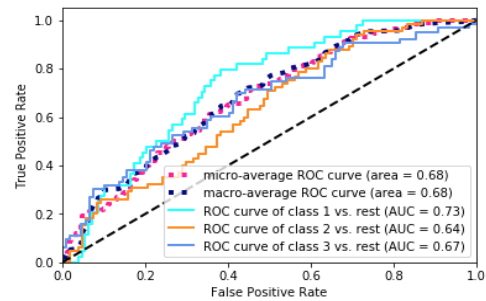

(g) Auto-Sklearn (101 nodes): all valid nodes for three groups

(h) Auto-Keras (101 nodes): all valid nodes for three groups

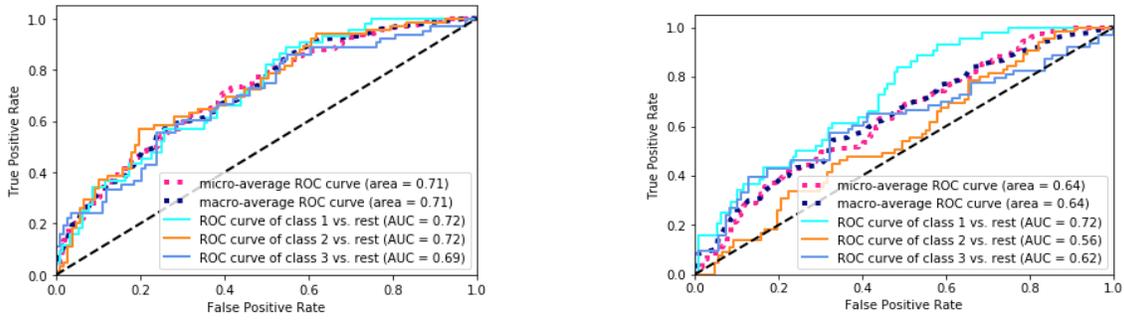

**Figure 3:** Auto-Sklearn/Auto-Keras classification results on 4 different settings with 22, 40, 77, and 101 nodes.
* class 1: ad, class 2: cn, class 3: mci

Figure 3 shows the results of Auto-Sklearn and Auto-Keras on the union of top-20, top-30, top-50 pivotal nodes, and all valid nodes for *setting 1* in terms of classification power on the testing data. The search budget (i.e., time in terms of both frameworks) is set to be the same for all experiments. The final models from Auto-Sklearn is typically the combination of dozens of models while those of Auto-Keras are deep convolutional neural networks (CNN) composed of a few hundred layers. Auto-Keras shows consistent lower performance, this might be related to that the sample size is too small for deep learning to generalize. Deep structure is meant for extracting meaningful local structures from large databases, in situations like ours, customized feature selection seems to be more powerful.

From Auto-Sklearn results shown in the left column of Figure 3, it can be seen that the union of top-20 pivotal nodes case has a relatively low micro/macro AUC when compared to the top-30 case. This means that the substructure with merely 22 nodes was not sufficient to provide strong discrimination between classes. However, increasing the threshold from 30 to 50 nodes and above (increasing the inclusion from 40 nodes to 77 nodes, and even include all 101 valid nodes) does not increase the performance. Thus, there is a clear diminishing return effect as more information is including in the model. The results indicate that 40 nodes (as the union of top-30 consistent ones from the weighted pairwise graph subtractions) provided the same discrimination power compared to the model with the entire valid node sets (101 nodes). In turn, this demonstrated the efficacy of our multi-view model for graph decomposition. Figure 4 below shows the detailed breakdown of one vs. the rest classification using the estimated probability for our identified most compact and discriminative subgraph structure (i.e., Union of top 30 pivotal nodes in ['am=(322/213)ad-mci', 'ac=(322/213)ad-cn','mc=mci-cn'] ). On separate ROC curves, we can optimize the cutoff point to obtain the confusion matrix below using Youden's J index method [19].

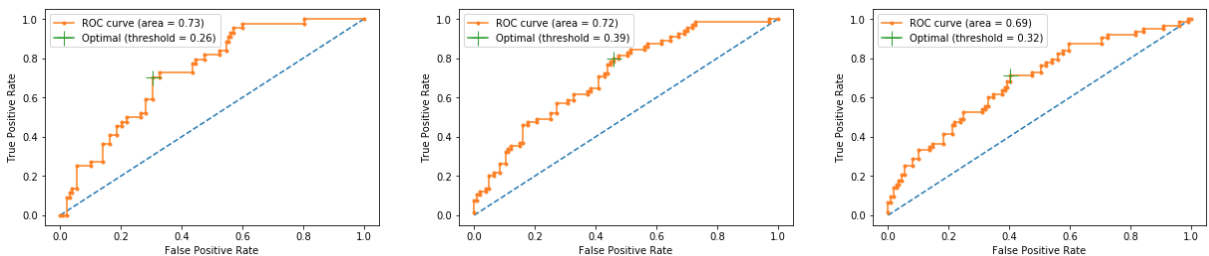

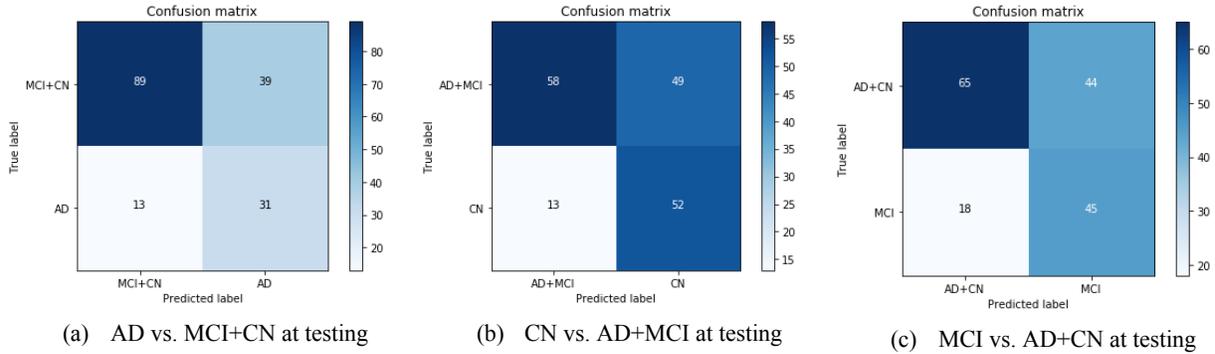

(a) AD vs. MCI+CN at testing    (b) CN vs. AD+MCI at testing    (c) MCI vs. AD+CN at testing

**Figure 4:** Detailed performance for Auto-Sklearn (40 nodes): Union of top 30 pivotal nodes in ['am=(322/213)ad-mci', 'ac=(322/213)ad-cn','mc=mci-cn']. Confusion matrices were determined by Youden's index.

In the next experiment (Figure 5), the discrimination based on the union of the pivotal nodes derived from graph subtractions (*setting 1*) with that of the union for the pivotal nodes from individual cohort (*setting 2*) was compared. Because the pivotal nodes for individual groups have a large degree of overlap, their union leads to much smaller sample size at the same selection threshold (i.e., top 30) and worse performance. When we increase the threshold to the top 50 nodes, the number of inclusion nodes become 44, getting close to that of the top 30 case for *setting 1* (40 nodes). Indeed, there is a large degree of overlapping of these 44 nodes with the 40 nodes in Figure 3(a). Further, experiments with randomly sampled nodes (40 and 60) indicated that the performance is worth than the model with the graph decomposition nodes.

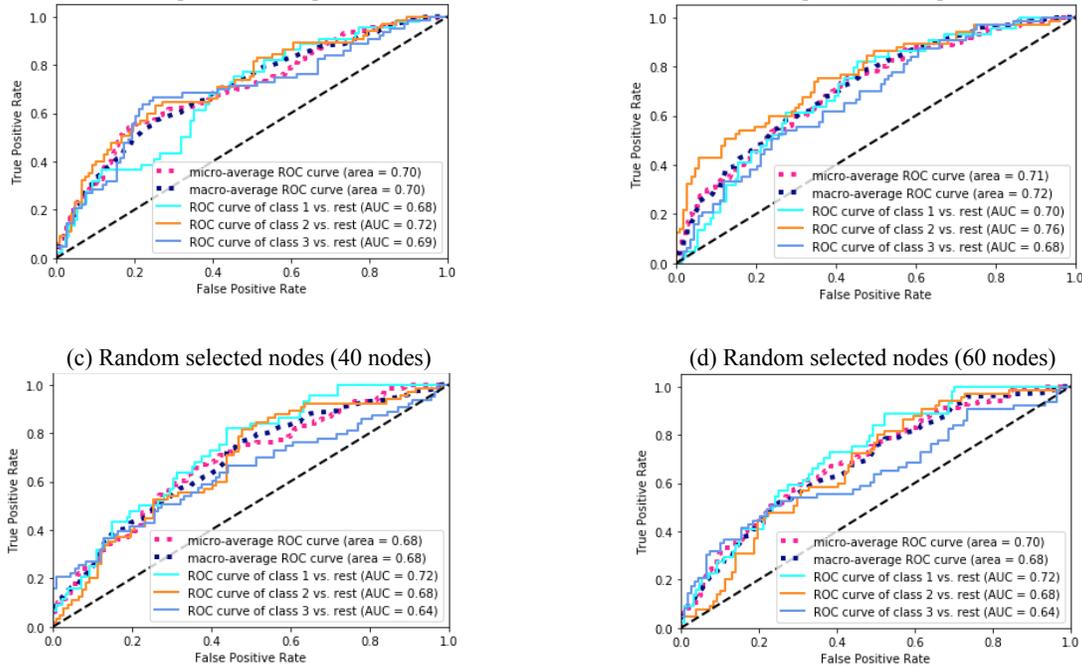

**Figure 5:** Discrimination based on the union of pivotal nodes of individual groups and random subgraphs

**Discovering discriminative markers from graph**

Besides evaluating the discrimination power of subgraphs, we were interested in visually exploring their graph morphology. Table 2 illustrates some of the distinguishing patterns from the AD vs. MCI, MCI vs. CN, and AD vs. CN experiments. In all three subfigures, the nodes displayed are subset from the table in the top left cell. We extracted networks for visualization from the raw differential brain imaging graph using symmetry cutoffs based on the values. This would help demonstrate strong association patterns. Interestingly, all the patterns picked by the model were concordant pairs (both regions grow or shrink at the same direction), and none of the discordant pairs were selected. For all three cases, the region "wm-lh-frontalpole" (white matter volume of the left hemisphere frontal lobe) represents the central of the structure. However, the differences between the graphs lie in the regions that the central node connects to. There are a large degree of overlap between AD vs. MCI and MCI vs. CN, showing that the mechanism of disease progression is consistent. The AD vs. CN has a much more compact graph structure and highly distinctive relationships. This difference seems to be aligned with the intuition that CN and AD are extreme groups in our cohort, and their difference should be more distinct. As a result, three out of the five relationships in CN vs. AD are unique ("wm-1h-frontalpole ←→ wm-lh-fusiform", "wm-1h-frontalpole ←→ wm-lh-fusiform", "wm-1h-frontalpole ←→ wm-lh-parahippocampal"), while AD vs. MCI only has one unique relationship ("wm-1h-frontalpole ←→ wm-lh-parsorbitalis") and MCI vs. CN also has one unique relationship ("wm-1h-frontalpole ←→ wm-lh-superiorparietal"). Overall, the left hemisphere's white matter volume was the dominant factor that distinguished the entire cohort in all three groups according to the graphs. Interestingly, although the candidate patterns are subgraphs associated with 40 pivotal nodes (including nodes associated with regions on the right hemisphere and corpus callosum), all of the strongest associations are on the left hemisphere.

## DISCUSSION AND CONCLUSION

The current study used MRI-based brain structural volumes to construct a novel graph-based approach for several AD-related classifications. In addition, newer methods that optimize the subgraph structure nodes were used to maximize algorithmic efficiency when searching through the large hyperparameter space. These methods were combined with longitudinal metrics that took into account how the size of different brain structures from an individual level changes over time. This encodes a diverse set of progression focused data to better understand AD.

**Table 2:** Illustration on critical patterns that distinguishes AD, MCI, and CN groups. Cutoffs were selected to make the graphs uncluttered for visualization purpose.

**Entities of interest**
**40 pivotal nodes identified in *setting 1* (top 30)**

| | | |
|---|---|---|
| 4th-Ventricle | Brain-Stem | Right-VentralDC |
| CC_Mid_Posterior | CC_Posterior | wm-lh-lateralorbitofrontal |
| Left-Cerebellum-Cortex | Left-Hippocampus | wm-lh-parsopercularis |
| Left-VentralDC | Right-Amygdala | wm-lh-precentral |
| Right-Putamen | Right-Thalamus-Proper | wm-lh-superiorparietal |
| wm-lh-fusiform | wm-lh-inferiortemporal | wm-rh-postcentral |
| wm-lh-middletemporal | wm-lh-parahippocampal | wm-lh-frontalpole |
| wm-lh-parstriangularis | wm-lh-postcentral | wm-lh-medialorbitofrontal |
| wm-lh-rostralmiddlefronta | wm-lh-superiorfrontal | wm-lh-parsorbitalis |
| wm-lh-supramarginal | wm-rh-inferiortemporal | wm-lh-precuneus |
| CC_Anterior | CC_Mid_Anterior | wm-lh-superiortemporal |
| Left-Amygdala | Left-Caudate | wm-rh-superiorfrontal |
| Left-Putamen | Left-Thalamus-Proper | |
| Right-Caudate | Right-Hippocampus | |

**AD vs. MCI [21 nodes total]**

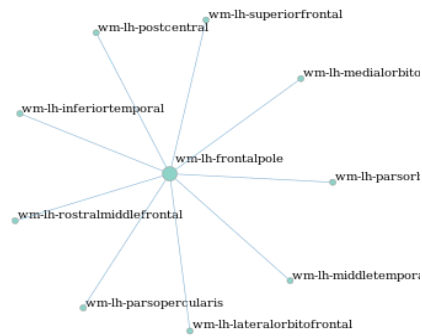

**MCI vs. CN [17 nodes total]**

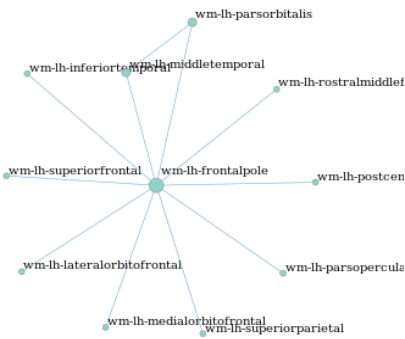

**CN vs. AD [15 nodes total]**

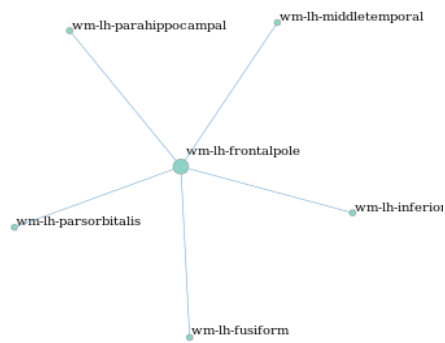

Many of the AD related computational studies use patient's imaging data cross-sectionally as it generally improves the subject numbers, and this tendency created several pre-processing pipelines. In this paper, the FreeSurfer

Longitudinal pipeline v 6.0 was used to take into account inter-subject differences to register each subject to an individualized common space. This reduces inter-subject variability that other cross-sectional pipelines lack. The structural volumes generated from this pipeline was then used to create a novel graph network primarily based on shrinkage or enlargement of different brain regions over time. One can think of this method as developing a representation of different regional structural changes with one another on a subject-level granularity. Once this matrix that encodes structural changes over time was created, the multi-view feature selection technique selected pivotal nodes important to the three classes of interest: AD, MCI, and CN. This subgraph structure with important nodes were the input to the subsequent classification experiments. These models were created using a series of very new ensembling methods that efficiently search through the vast hyperparameter space. Auto-Sklearn and Auto-Keras were employed to train and evaluate. The results indicated that a substructure constructed by our graph decomposition method is sufficient to provide the discrimination power. All of the combinations of One vs. All experiments were also completed. The AD vs. MCI + CN and CN vs. AD + MCI tests had AUCs of 0.73 and 0.72, respectively. Interestingly, the MCI vs. AD + CN had a comparable AUC score of 0.69. This classification scheme is not typically used in current studies, but this combination may be important for understanding what structures are changing within those who are at risk of development AD. After classification, the critical graph nodes that drove the models' predictions were visualized. The results also indicated that the left hemisphere determines progression. The white matter volume of the left hemisphere frontal pole was considered the central node in all of the phenotype classifications. Studies have shown that white matter changes occured in frontal, parietal, and temporal lobes in addition to hippocampal regions which are present in the subgraphs [20,21]. Further, a longitudinal DTI based study found that the medial temporal lobe underwent extensive changes in their cohort [4], and this brain region is seen in the above figures. The structural difference between different phenotypes is also interesting. The AD vs. CN experiment had the least amount of nodes which could mean that only a few regions that change significantly are needed to classify between these extreme groups. The MCI vs. CN graph contained more nodes which backs this hypothesis as it make take more less significant changes for proper classification. Further, the AD vs. MCI classification is most likely the hardest classification problem, and this experiment led to the most nodes in the subgraph. The authors used only a small portion of the available structural brain volumes available from the FreeSurfer pipeline. There could in fact be more discriminative features that were overlooked. Another potential limitation would be the differing criteria to diagnose AD. These data were taken from multiple studies within ADNI, and the inherent heterogeneity in these samples may introduce bias into the results. Along the same lines, the dataset could have been expanded if samples were taken cross-sectionally instead of longitudinally. Lastly, the samples were not adjusted using demographics like age and ethnicity.

**Acknowledgement:** Data collection and sharing for this project was funded by the Alzheimer's Disease Neuroimaging Initiative(ADNI) (National Institutes of Health Grant U01 AG024904) and DOD ADNI (Department of Defense award number W81XWH-12-2-0012). ADNI is funded by the National Institute on Aging, the National Institute of Biomedical Imaging and Bioengineering, and through generous contributions from the following: AbbVie, Alzheimer's Association; Alzheimer's Drug Discovery Foundation; Araclon Biotech; BioClinica, Inc.; Biogen; Bristol-Myers Squibb Company; CereSpir, Inc.; Cogstate; Eisai Inc.; Elan Pharmaceuticals, Inc.; Eli Lilly and Company; EuroImmun; F. Hoffmann-La Roche Ltd and its affiliated company Genentech, Inc.; Fujirebio; GE Healthcare; IXICO Ltd.; Janssen Alzheimer Immunotherapy Research & Development, LLC.; Johnson & Johnson Pharmaceutical Research & Development LLC.; Lumosity; Lundbeck; Merck & Co., Inc.; Meso Scale Diagnostics, LLC.; NeuroRx Research; Neurotrack Technologies; Novartis Pharmaceuticals Corporation; Pfizer Inc.; Piramal Imaging; Servier; Takeda Pharmaceutical Company; and Transition Therapeutics. The Canadian Institutes of Health Research is providing funds to support ADNI clinical sites in Canada. Private sector contributions are facilitated by the Foundation for the National Institutes of Health (www.fnih.org). The grantee organization is the Northern California Institute for Research and Education, and the study is coordinated by the Alzheimer's Therapeutic Research Institute at the University of Southern California. ADNI data are disseminated by the Laboratory for Neuro Imaging at the University of Southern California.

**REFERENCES**
1. Aisen PS, Cummings J, Jack CR Jr, Morris JC, Sperling R, Frölich L, Jones RW, Dowsett SA, Matthews BR, Raskin J, Scheltens P, Dubois B. On the path to 2025: understanding the Alzheimer's disease continuum. Alzheimers Res Ther 2017 Aug 9;9(1):60. PMID:28793924
2. Huang M, Yang W, Feng Q, Chen W, Alzheimer's Disease Neuroimaging Initiative. Longitudinal measurement and hierarchical classification framework for the prediction of Alzheimer's disease. Sci Rep 2017 Jan 12;7:39880.

3. Spasov S, Passamonti L, Duggento A, Liò P, Toschi N, Alzheimer's Disease Neuroimaging Initiative. A parameter-efficient deep learning approach to predict conversion from mild cognitive impairment to Alzheimer's disease. Neuroimage 2019 Jan 14;189:276–287. PMID:30654174
4. Mayo CD, Mazerolle EL, Ritchie L, Fisk JD, Gawryluk JR, Alzheimer's Disease Neuroimaging Initiative. Longitudinal changes in microstructural white matter metrics in Alzheimer's disease. Neuroimage Clin 2017;13:330–338.
5. Rathore S, Habes M, Iftikhar MA, Shacklett A, Davatzikos C. A review on neuroimaging-based classification studies and associated feature extraction methods for Alzheimer's disease and its prodromal stages. Neuroimage 2017 Jul 15;155:530–548. PMID:28414186
6. Zhou T, Thung K-H, Zhu X, Shen D. Feature Learning and Fusion of Multimodality Neuroimaging and Genetic Data for Multi-status Dementia Diagnosis. Machine Learning in Medical Imaging Springer International Publishing; 2017. p. 132–140.
7. Schouten TM, Koini M, Vos F de, Seiler S, Rooij M de, Lechner A, Schmidt R, van den Heuvel M, Grond J van der, Rombouts SARB. Individual classification of Alzheimer's disease with diffusion magnetic resonance imaging. Neuroimage 2017 May 15;152:476–481. PMID:28315741
8. Reid AT, Evans AC. Structural networks in Alzheimer's disease. Eur Neuropsychopharmacol 2013 Jan;23(1):63–77.
9. Pereira JB, van Westen D, Stomrud E, Strandberg TO, Volpe G, Westman E, Hansson O. Abnormal Structural Brain Connectome in Individuals with Preclinical Alzheimer's Disease. Cereb Cortex 2018 Oct 1;28(10):3638–3649.
10. John M, Ikuta T, Ferbinteanu J. Graph analysis of structural brain networks in Alzheimer's disease: beyond small world properties [Internet]. Brain Structure and Function. 2017. p. 923–942. [doi: 10.1007/s00429-016-1255-4]
11. Phillips DJ, McGlaughlin A, Ruth D, Jager LR, Soldan A, Alzheimer's Disease Neuroimaging Initiative. Graph theoretic analysis of structural connectivity across the spectrum of Alzheimer's disease: The importance of graph creation methods. Neuroimage Clin 2015 Jan 13;7:377–390. PMID:25984446
12. Longitudinal Processing - Free Surfer Wiki [Internet]. [cited 2019 Mar 10]. Available from: https://surfer.nmr.mgh.harvard.edu/fswiki/LongitudinalProcessing
13. Reuter M, Schmansky NJ, Rosas HD, Fischl B. Within-subject template estimation for unbiased longitudinal image analysis. Neuroimage 2012 Jul 16;61(4):1402–1418. PMID:22430496
14. Feurer M, Klein A, Eggensperger K, Springenberg J, Blum M, Hutter F. Efficient and Robust Automated Machine Learning. In: Cortes C, Lawrence ND, Lee DD, Sugiyama M, Garnett R, editors. Advances in Neural Information Processing Systems 28 Curran Associates, Inc.; 2015. p. 2962–2970.
15. Jin H, Song Q, Hu X. Auto-Keras: Efficient Neural Architecture Search with Network Morphism [Internet]. arXiv [csLG]. 2018. Available from: http://arxiv.org/abs/1806.10282
16. Hutter F, Hoos HH, Leyton-Brown K. Sequential Model-Based Optimization for General Algorithm Configuration (extended version). Available from: https://www.cs.ubc.ca/~hutter/papers/10-TR-SMAC.pdf
17. Hida T, Hitsuda M. Gaussian Processes. American Mathematical Soc.; ISBN:9780821887639
18. ADNI | Alzheimer's Disease Neuroimaging Initiative. [cited 2019 Mar 10]. Available from: http://adni.loni.usc.edu/
19. Youden WJ. Index for rating diagnostic tests. Cancer 1950 Jan;3(1):32–35. PMID:15405679
20. Acosta-Cabronero J, Nestor PJ. Diffusion tensor imaging in Alzheimer's disease: insights into the limbic-diencephalic network and methodological considerations. Front Aging Neurosci 2014 Oct 2;6:266. PMID:25324775
21. Sexton CE, Kalu UG, Filippini N, Mackay CE, Ebmeier KP. A meta-analysis of diffusion tensor imaging in mild cognitive impairment and Alzheimer's disease. Neurobiol Aging 2011 Dec;32(12):2322.e5–18. PMID:20619504